\documentstyle[11pt,epsf]{article} 
\begin{document}
\centerline{\bf SPECTRAL ENERGY DISTRIBUTIONS OF GAMMA-RAY BLAZARS}
\vskip 0.5 true cm
\centerline{\bf Laura Maraschi }
\centerline{\it Osservatorio Astronomico di Brera, via Brera 28, 20121 Milano,
 Italy} 
\centerline{\bf Giovanni Fossati}
\centerline{\it Scuola Internazionale Superiore di Studi Avanzati, via Beirut
2--4, 34014 Trieste, Italy} 

\vskip 0.5 true cm
\noindent 
{ABSTRACT }

\noindent
Average Spectral Energy Distributions (SED) for different subgroups of 
blazars are derived from available homogeneous (but small) data sets, 
including the gamma--ray band.
Comparing Flat Spectrum Radio Quasars (FSRQ) with BL Lacs extracted
from radio (RBL) or X--ray surveys (XBL) remarkable differences and 
similarities are apparent: i) in all cases the overall SED from radio 
to gamma--rays shows two peaks; ii) the first and second peak fall in different 
frequency ranges for different objects, with a tendency for the most
luminous objects to peak at lower frequencies; iii) the ratio between the
two peak frequencies {\it seems to be constant}, while the luminosity ratio
between the high and low frequency component increases from XBL to RBL and
 FSRQ. The variability 
properties, (amplitude and frequency dependence) are similar in different
objects if referred to their respective peak frequencies.
Finally, comparing spectral snapshots obtained at different epochs, 
the intensities of the two components at frequencies close 
and above their respective peaks seem to be correlated. The relevance
of these properties for theoretical models is briefly discussed.

\vskip 0.2 true cm
\noindent
{1. THE BLAZAR FAMILY }
\vskip 0.1 true cm

\noindent
In 1978 Ed Spiegel proposed the name "blazar" to designate a class of objects 
including BL Lacs {\it as well as } flat spectrum, radio loud quasars 
(Angel \& Stockman, 1980). Although  there are differences
among members of the class, the underlying hypothesis was that at least the 
phenomena associated with the continuum  derive from a common process,
most likely a jet of relativistically moving plasma.

The opening of the high energy windows of X--ray and gamma--ray astronomy 
revealed further differences within the blazar family.
X--ray astronomy led to the discovery of a new subclass, the so called
X--ray selected BL Lacs, differing from the classical radio-selected
BL Lacs in the relative strength of their X--ray and radio emissions
and in a lesser degree of "activity" in the radio to optical emission
(e.g. Urry and Padovani, 1995).
 
More recently, the Compton Gamma Ray Observatory revealed
that in many cases a substantial fraction and in some cases the bulk of the 
emitted power is released in this very high energy band. 
This fundamental discovery with far reaching theoretical implications 
extensively discussed at this meeting,
poses the problem  as to why some blazars are detected in gamma--rays
and some are not, or in other words whether there are significant
differences in the gamma--ray emission of different blazars.

In the following we will discuss the observed spectral energy distributions 
of blazars from radio to gamma--rays, and their variability, showing that 
there are clear systematic
differences between the continua of different blazar subclasses.
Nevertheless a unitary approach is still possible, maintaining
a common  process of energy release in a relativistic jet and  relating the
observed differences to different physical conditions in or around the jet.

\vskip 0.2 truecm
\noindent
{2. SPECTRAL ENERGY DISTRIBUTIONS OF BLAZARS}
\vskip 0.1 truecm

\noindent
Average spectral energy distributions of blazars have been constructed
recently, taking advantage of the fact that the complete samples of the
1 Jy BL Lacs (RBLs), the EMSS BL Lacs (XBLs) and a small but complete sample 
of FSRQs (Brunner et al. 1994) have all been observed with the {\it ROSAT} 
PSPC allowing to derive "uniformly"  
X--ray fluxes and in most cases spectral shapes in the 0.1 -- 2 keV
range (Sambruna et al. 1996a, and references therein). 
Average fluxes at radio, mm (230 GHz), IR and optical frequencies 
were taken from the literature.

The SEDs of each class, plotted in a  $\nu F(\nu)$ vs. $\nu$ diagram,
are smoothly convex  up to the X--ray range  and  show a broad peak (Fig.~1).
The peak frequency increases from FSRQs to RBLs and XBLs. The X--ray spectrum
follows the general trend of continuous steepening only
in the case of XBLs, while for RBLs and FSRQs the X--ray spectrum is flatter
than the optical to X--ray extrapolation, suggesting a new spectral
component. If RBLs are splitted into three
 subgroups according to the shape of their optical to X--ray continua
(convex, straight or concave) the SEDs of the three subgroups at  
lower frequencies follow the same patterns as the three main groups,
suggesting that we have to do with a continuous spectral sequence 
within the blazar family, rather than with separate spectral classes.

\begin{figure}[t]
\epsfxsize=12 truecm
\centerline{\epsffile[18 204 592 630]{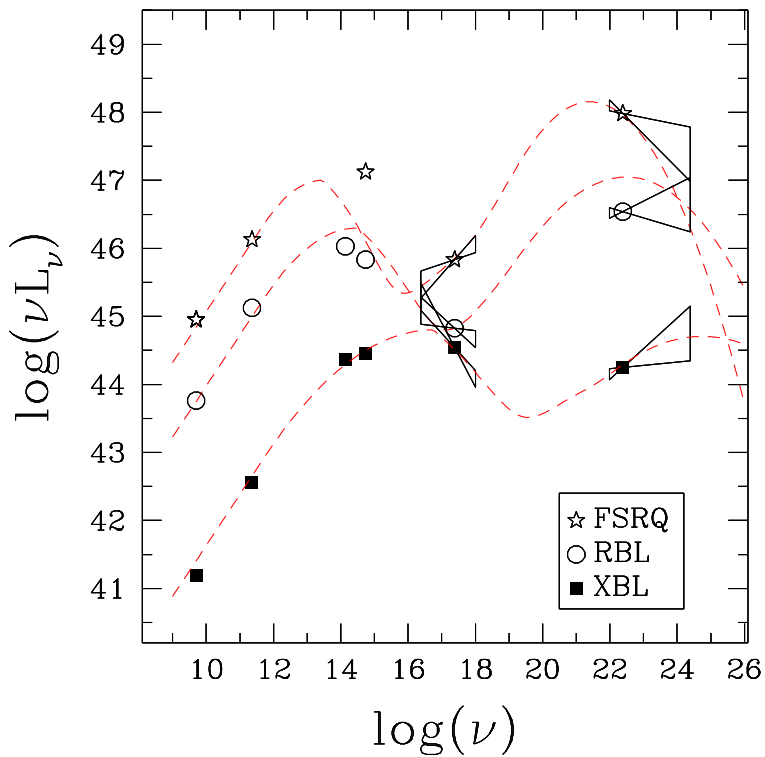}}
\caption{\sf{Spectral Energy Distributions of blazar subclasses. The points 
up to the X--ray range derive from average fluxes for complete samples
of blazars (Sambruna et al. 1996a). The average slopes in the X--ray range
are also shown. Average gamma--ray intensities and spectral shapes
were obtained from results by Comastri et al. (1997) as explained in the text.
The dashed lines are analytic approximations to the SEDs meant to
merely "describe" the observed trends. }
}
\end{figure}

Unfortunately the gamma--ray observations are at present less "systematic".
In order to add average gamma--ray fluxes and spectral indices to the 
SEDs for XBLs, RBLs and FSRQs we will use the correlation between gamma--ray 
properties and X--ray properties found by Comastri et al. (1997)
 for 37 FSRQs, 11 RBLs and 5 XBLs with measured fluxes in both
bands (see also Ghisellini this volume). For the joint 
sample, the gamma to X--ray flux ratio is strongly correlated
with the radio to optical flux ratio, characterized by the two point
spectral index $\alpha_{RO}$. Based on the above correlation we assign a
gamma--ray flux to each subclass according to the average $\alpha_{RO}$
parameter of the class, the average X--ray flux for the class and the
gamma to X--ray flux ratio defined by the correlation of Comastri et al. 
(1997). In addition we plot for each class the average spectral
 shape in the 0.1 -- 5 GeV band 
measured by the EGRET experiment on board CGRO for the above objects.

The results are shown in Fig.~1, admittedly very preliminary, but 
highly suggestive. Clearly a second peak is present in the SED of
each subgroup in a very broad gamma--ray range. The location of this
second peak can  be guessed from the shape of the gamma ray spectrum.
Depending on whether the gamma--ray spectrum is steep , medium or flat
($\alpha_{\gamma} > 1, \simeq 1, < 1$), the peak 
must fall below within or above the EGRET range.

The dashed curves in Fig.~1 are drawn mainly to guide the eye.
However we mention that the low frequency part is a parameterization
discussed in Fossati et al. (1996) based on the hypothesis that the
peak frequency of the first spectral component is inversely related 
to the luminosity in this component ($\nu_{peak} \propto L^{-1.5}$ in
this case). The only point deviating strongly from the analytical
description is the optical flux of FSRQ. While this discrepancy could be 
real, we note that for the 8 objects in the sample, the adopted magnitudes
 probably have large errors, since these objects were scarcely observed
and the magnitudes used were mostly retrieved from the
V\'eron-Cetty \& V\'eron catalog (V\'eron-Cetty \& V\'eron, 1993). 

The lines describing the second
spectral component, peaking at very high frequencies have been 
drawn assuming that the second peak frequency has a fixed ratio ($10^8$
to the first one and a height proportional to the average 
radio luminosity of each subclass.
It is remarkable how well the average X--ray spectra fit into this
very simple scheme. 

Summarizing: the SEDs of blazars show two broad peaks. The first
can fall between $10^{14} - 10^{17}$ Hz, the second between
$10^{22} - 10^{25}$ Hz depending on the object.
However for a given object the two peak frequencies are correlated,
the present data being compatible with a constant ratio. 
The strength of the gamma--ray peak with respect to the lower frequency
one increases with increasing $\alpha_{RO}$. 
 
\vskip 0.2 true cm
\noindent 
{ 3. VARIABILITY AND SPECTRAL ENERGY DISTRIBUTIONS }
\vskip 0.1 true cm

\noindent
The above properties can be  verified in several, diverse, prototype objects
for which also variability data have been obtained, namely 3C 279, 
a weak lined FSRQ (Maraschi et al. 1994), Mrk 421, a typical XBL 
(Macomb et al. 1995, 1996), and PKS 0528+134, a typical strong lined 
FSRQ (Sambruna et al. 1996b). We can summarize the comparison of broad band 
spectral snapshots obtained at different epochs for each object, stating 
that the variability is similar in all these objects, once  the different 
peak frequencies  of their spectral components are taken into account.
More explicitly, in all cases
the variations are larger and the spectra tend to be harder in higher intensity
states {\it at frequencies larger than the peak frequencies of the two
spectral components}. 
For instance a hardening is seen clearly in the IR to UV continuum of 3C 279
($\nu_{peak,1}$  falls in the FIR) 
and in the X--ray spectrum of Mrk 421 ( $\nu_{peak,1}$ falls in the 
soft X--ray band). 
In Mrk 421 a hardening is observed also in the high energy spectrum, 
where the TeV emission varies more than the GeV emission ($\nu_{peak,2} \simeq
100$ GeV.
In PKS 0528 +134, the gamma--ray slope measured by EGRET is $> 1$ 
($\nu_{peak,2} \simeq 100$ MeV) and flattens significantly at high intensity.
It is remarkable that a shift by three decades to lower frequencies of the 
Mrk 421 data yields a very good match with the high and low gamma--ray 
spectra of PKS 0528+134.


Finally it is important to stress that as long as one considers the energy
ranges above each peak (at $\nu_{peak,1}$ and $\nu_{peak,2}$), where
the largest variability is observed, the intensities of the high
and low energy branch in the same object vary in a correlated fashion.
We are not aware of any counterexample to this behaviour.

\vskip 0.2 true cm
\noindent
{4. GENERAL THEORETICAL CONSIDERATIONS}
\vskip 0.1 true cm

\noindent
The above properties strongly support models involving synchrotron and Inverse
Compton radiation in a relativistic jet to account for the two components 
in the SEDs. 
The arguments are i) that the spectral shapes of the two components 
are "similar" as expected for synchrotron and IC radiation from the same 
relativistic electrons (except for effects of self--absorption); ii) the two 
components vary in a correlated fashion. On the other hand  present 
observations are insufficient to decide whether the seed photons are the 
synchrotron photons themselves (SSC) or photons present outside the jet 
(External Compton, EC) (see for a review Sikora 1994, and refs. therein). 

Comparative spectral fits with both models to the same data and a discussion
of the expected correlations in the two cases can be found in 
Ghisellini et al. (1996) and Maraschi and Ghisellini (1996).
Briefly, if the variations are due to changes in the relativistic
electron spectrum, the SSC model or the ``mirror model'' of Ghisellini 
and Madau (1996) predict that  the Compton branch varies more than the 
corresponding synchrotron branch, while in the EC model the variations 
should be proportionate. However if the bulk Lorentz factor of the emitting 
region changes, the EC model variability can mimic the SSC one, and 
intermediate cases are possible if other parameters vary as well.

Short time scale variability at different frequencies (i.e. multifrequency 
light curves as obtained thus far for the BL Lac object PKS 2155--304,
observed simultaneously with ASCA, EUVE and IUE, Urry et al. 1996;
and for Mrk 421 observed simultaneously by the Whipple observatory, ASCA, EUVE 
and from ground, Buckley et al. 1996) offers the best means to discriminate 
between models. 
However the problem is complicated in practice 
by the necessity of taking into account the "structure" of the source,
as indicated by the "delays" observed in the above sources.
This is a new territory which needs to be explored (see the papers by
Mastichiadis and Ghisellini in this volume).

A final comment concerns  the trends in the
SEDs shown in Fig.~1. The astrophysical cause of this trend is
presently unknown. Ideas under discussion involve
self--limited acceleration in the presence of strong radiative losses
(Fossati et al. 1996) or different external (or intrinsic) physical
conditions for the jet development (Sambruna et al. 1996b).

It would be extremely important to determine the physical parameters 
of the emitting region. If the SSC model applies to the three subclasses,
then the constant ratio between the two peak frequencies implies
that the energy of the relativistic electrons radiating at the peak
$(\gamma_e)^{peak}$ is the same in all objects. Consequently the magnetic
field should be lower in objects with higher luminosity, peaking at lower
frequencies (i.e. FSRQ).

If the spectral sequence correponds to the increasing dominance
of the external radiation field (with constant, typical, rest frame frequency
$\simeq 10^{15}$ Hz), then, in order to maintain a constant
ratio between the observed peak frequencies, $(\gamma_e)^{peak}$ 
should be lower  in FSRQs and the magnetic field should be constant
for the three subclasses.

\vskip 0.2 true cm
\noindent
{\bf References}
\parindent=0 pt
\everypar={\hangindent=2.6pc}
\vskip 0.1 true cm

Angel, J.R.P., \& Stockman, H.S. 1980, ARA\&A, 18, 321 

Buckley, J.H., et al. 1996, ApJ, in press

Brunner, H., Lamer, G., Worral, D.M., \&  Staubert, R., 1994, A\&A, 287, 43

Comastri, A., Fossati, G., Ghisellini, G., \& Molendi, S., 1997, ApJ, submitted

Fossati, G., Celotti, A., Ghisellini, G., \& Maraschi L., 1996 M.N.R.A.S., submitted

Ghisellini, G., \& Madau, P. 1996, M.N.R.A.S., 280, 67 

Ghisellini, G., Maraschi, L., \& Dondi, L., 1996, A\&A in press. 
 
Macomb, D.J., et al. 1995, ApJ, 449, L99

Macomb, D.J., et al. 1996, ApJ, 459, L111

Maraschi, L., et al. 1994, ApJ, 435, L91

Maraschi, L., \& Ghisellini, G. 1996, Proc. of "Variability of Blazars", Miami
Febr. 1996.

Sambruna, R.M., Maraschi, L., \&  Urry, C. M. 1996a, ApJ, 463, 444

Sambruna, R.M., et al. 1996b, ApJ, in press 

Sikora, M. 1994, ApJS, 90, 923

Urry, C.M., \& Padovani, P. 1995, PASP, 107, 803

Urry, C.M., et al. 1996, ApJ, submitted

Veron--Cetty, M.P. and Veron, P., 1993, ESO Scientific Report No. 13

\end{document}